# Facilitating the systematic nanoscale study of battery materials by atom probe tomography through *in-situ* metal coating

Mahander P Singh[1], Eric V Woods[1], SeHo Kim[1,2], Chanwon Jung[1], Leonardo S. Aota[1], Baptiste Gault[1,3]

1. Max-Planck-Institut für Eisenforschung GmbH, Max-Planck-Straße 1, 40237 Düsseldorf, Germany
2. *now at* Department of Materials Science and Engineering, Korea University, Seoul 02841, Republic of Korea
3. Department of Materials, Royal School of Mines, Imperial College London, Prince Consort Road, London SW7 2BP, UK

Corresponding authors e-mail address: (MPS) m.singh@mpie.de, (B.G) b.gault@mpie.de

## Abstract

Through its capability for 3D mapping of Li at the nanoscale, atom probe tomography (APT) is poised to play a key role in understanding the microstructural degradation of lithium-ion batteries (LIB) during successive charge- and discharge cycles. However, APT application to materials for LIB is plagued by the field induced delithiation (deintercalation) of Li-ions during the analysis itself that prevents the precise assessment of the Li distribution. Here, we showcase how a thin Cr-coating, *in-situ* formed on APT specimens of NMC811 in the focused-ion beam (FIB), preserves the sample's integrity and circumvent this deleterious delithiation. Cr-coated specimens demonstrated remarkable improvements in data quality and virtually eliminated premature specimen failures, allowing for more precise measurements via. improved statistics. Through improved data analysis, we reveal substantial cation fluctuations in commercial grade NMC811, including complete grains of LiMnO. The current methodology stands out for its simplicity and cost-effectiveness and is a viable approach to prepare battery cathodes and anodes for systematic APT studies.

## 1 Introduction

Li-ion batteries (LIB) appears to be among the most viable and scalable energy storage technologies, that are necessary to meet the growing demands of future green energy needs [1]. In the realm of battery material research, cathodes garner significant attention, primarily due to their relatively lower gravimetric energy densities when compared to commercially available anodes like graphite or pure lithium[2,3]. Achieving enhanced performance in battery cathodes materials requires a deep understanding of their atomic-scale structure and composition [4–8]. Understanding the atomic-scale changes in battery materials is complicated by the small size and high mobility of Li ions and the intricate compositions of complex oxides. This represents a significant challenge for all characterization techniques, particularly when mapping the distribution of light elements like Li and O within a matrix of heavier elements such as Ni, Co and Mn. For example, X-ray and electron microscopy techniques, which are the most versatile and reliable material analysis methods, cannot be used for the direct observation of Li distributions due e.g. weak atomic scattering, low radiation energy and severe sensitivity to beam damage [8–10].

Atom probe tomography (APT) [11] has sub-nanometer precision and an elemental sensitivity not related to the atomic weight of the analyzed species has the potential for mapping Li along with the other elements [5,12]. However, APT is underpinned by an intense electric field that can drive preferential lithium outward migration and cause *in-situ* delithiation [13,14]. Battery materials are also often reactive oxides, that require careful handling and sample transfer [15]. There are only few studies of battery materials using APT, as recently reviewed by Li et al. [16] and more worryingly maybe, there are discrepancies between results, owing to these issues [17–20]. It is also important to acknowledge that compositional inhomogeneity at the micro or nano-scale, particularly in bulk prepared samples, can also cause such variations in the results [21]. These inhomogeneities can introduce inconsistencies in battery performance, and it is imperative to address them. Probing such inhomogeneities using APT can further pave the way to understand batteries better.

To address this, Kim et al. [12], demonstrated that delithiation can be circumvented by a conducting layer deposited on the APT specimen's surface. They deposited a thin layer of Ni using physical vapor deposition (PVD) onto sharpened APT specimens. A similar approach had been used to study the microstructure of the coatings themselves or the interface between the specimen and the coating [22–24]. Coating have also been used for improved analytical performance, as reported by e.g. Larson et al. [25] and Seol et al. [26], who demonstrated improvement in mass resolving power through reduction in thermal tails from laser pulsing. In another study, Kim et al [27] embedded $BaTiO_3$, a perovskite-structured material, into a nickel matrix. Application of the metallic coating effectively prevented charge penetration caused by the electrostatic field and enabled the APT analysis.

While the application of thin metal coatings onto APT specimens appears promising, it is important to acknowledge that these specimens needed to be exposed to ambient atmosphere during the transfer to the deposition chamber [12,25,26]. This exposure can be problematic for reactive samples, including battery materials. Applying a thin, conformal and continuous film on the surface of sharp needle-shaped APT specimens can also be a challenge. Recently, our group proposed a simple procedure for applying thin metal coatings directly within the scanning-electron microscope - focused-ion beam (SEM-FIB) used to fabricate the needle-shaped specimens for APT [28], inspired by previous reports [29,30].

In the present work, we analyze by APT commercial NMC811 ($LiNi_{0.8}Co_{0.1}Mn_{0.1}O_2$) powder samples using the Cr-coated specimens by using this *in-situ* deposition technique. The layered NMC ($LiNi_xMn_yCo_zO_2$) cathodes are the most widely investigated commercial cathode materials for Li- batteries, owing to it high capacity and good thermal stability [31]. Cr was selected for this purpose because of good adhesion and similarity of the evaporation fields. We develop a comprehensive understanding of how the coating impacts the APT data quality for this critical battery materials, laying the ground for systematic studies that allow us to conclude that within a batch of commercial NMC811 powder, there are significant variations in the particle's composition, including some Mn-rich (Li-Mn-O) particles.

## 2 Materials and methods

Commercial NMC811 powder with a stoichiometry $LiNi_{0.8}Co_{0.1}Mn_{0.1}O_2$ was sourced from Tragray. The nominal composition of the NMC811 ($LiNi_{0.8}Co_{0.1}Mn_{0.1}O_2$) is 25 Li, 20 Ni, 2.5 Co, 2.5 Mn and 50 O in at.%, as suggested by the supplier, which was used as a baseline for the compositions measured by APT below. To reduce the atmospheric interaction, a $N_2$-filled glovebox from the Laplace project was used [32]. The NMC811 powder was evenly dispersed onto a Cu-tape, which was placed on a flat stub and subsequently transferred to the dual beam SEM-FIB Thermo Scientific Helios 5 instrument. Sample transfer was done using the Ferrovac vacuum suitcase at room temperature. The detailed procedure of the sample preparation is available in Ref. [28]. APT specimens were Cr-coated using the *in-situ* FIB sputtering workflow detailed below. $LiMn_2O_4$ (LMO) powder (spinel structure) were sourced from Thermoscientific and prepared using the same protocol.

APT analyses were conducted using both a LEAP 5000 XR and a LEAP 5000 XS (CAMECA[TM]) Instruments Inc. Madison, WI, USA, operating in UV-laser pulsing mode within an ultrahigh vacuum environment ($10^{-11}$ mbar). Data was collected at a base temperature of 60 K, with a pulse repetition rate of 100 kHz and a detection rate of 0.5 ions per 1000 pulse on average. Laser pulse energy was varied during the study, as detailed below. Data analysis was done using the APSuite 6.1 software.

Cr-coated APT specimens were analyzed using a transmission electron microscope (TEM) (FEI Titan Themis microscope with an image corrector, operated at 300 kV). Additionally, energy - dispersive X-ray (EDX) analyses and high-angle annular dark field (HAADF) imaging were conducted in scanning transmission electron microscopy (STEM) mode.

## 3 Results and discussion

### 3.1 Delithiation

To illustrate the uncontrolled analysis conditions for LIB materials by APT, we repeated experiments from Ref. [12]. Preliminary data was acquired by transferring specimen to the atom probe in an ultrahigh vacuum (<$10^{-8}$ mbar) using the suitcase. Most specimens faced premature fracture during analysis. However, a limited number of APT specimens ran successfully, and figure 1a-d summarizes data acquired under varying laser pulse energy of 5 pJ, 10 pJ, 15 pJ and 20 pJ, respectively. In each case, the top image illustrates the detector event maps, i.e. a histogram of the detector hit density, followed by 3D-reconstruction along both the X-Z and the X-Y planes. As the laser pulse energy increases, the heterogeneity in the ion field evaporation is markedly more apparent. In the 3D-reconstrcution of the 5pJ section, shown in figure 1a, only Ni ions are displayed, with only few Li-ions detected. Conversely, at 20 pJ, one can exclusively observe the evaporation of Li-ions. These experiments highlight that avoiding delithiation involves balancing both the electrostatic field conditions and laser power.

### 3.2 Coating

In an endeavor to address these challenges, in situ Cr-coating was employed on the freshly prepared APT specimens of NMC811 samples. The same procedure for APT specimen preparation was used, with Cr-coating deposited based on the protocol described in Reference [28]. To probe the adherence and estimate the extent of the Cr-coating, APT specimens were also prepared on a Mo grid using a correlative holder, as described in Reference [33]. For TEM studies sharpened specimens were only coated from a single side with 30 kV, 80 pA current for 30 sec. Figure 2a and b shows the electron and ion beam image of the specimen placed along the Cr-lamella in FIB.

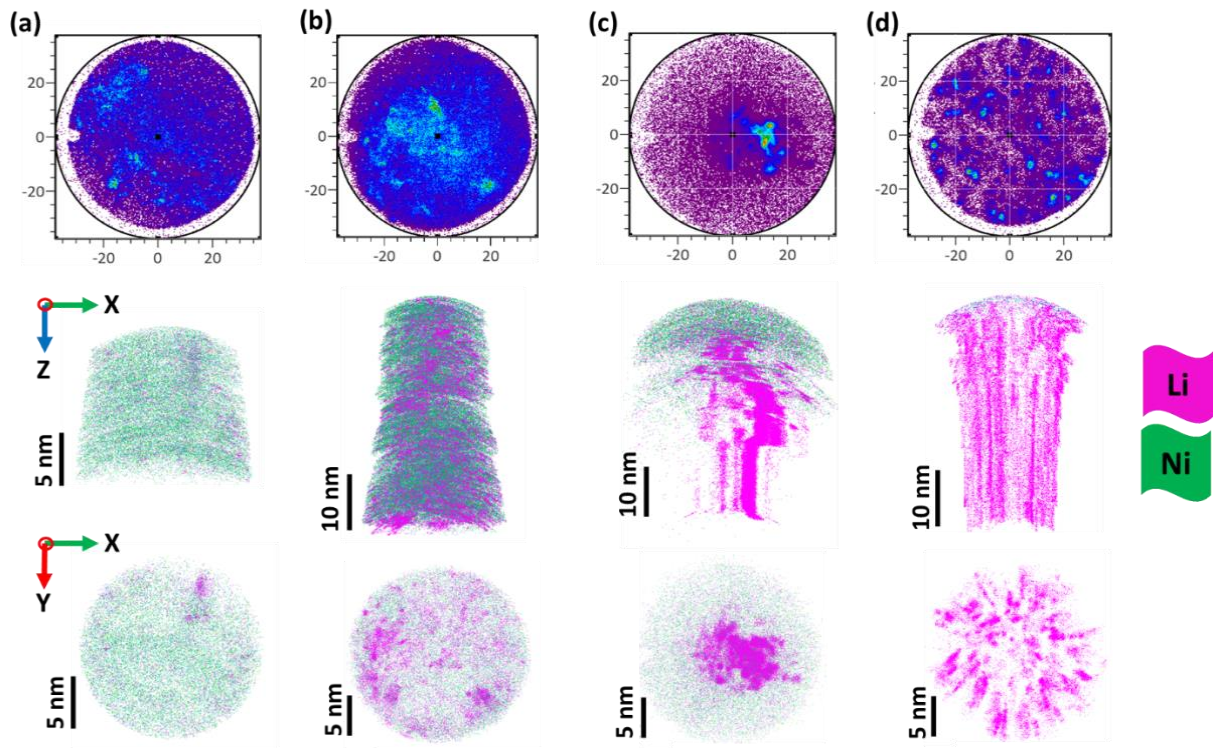

**Figure 1.** *Detector event maps obtained for NMC811 samples, evaporated at (a) 5 pJ, (b) 10 pJ, (c) 15 pJ, and (d) 20 pJ of laser energies. 3D-reconstructions of the measurements performed under respective laser energies is provided below in both XZ plane and XY plane.*

Figure 2c and d shows the STEM-HAADF and STEM-EDS of the coated sample, respectively. A thin bright line delineates the outer shell of Cr-coating from the battery region in the HAADF image. The EDS-map shows a continuous, thin and conformal Cr-coating along the whole length of the specimen. Specimens imaged by TEM could not be successfully analyzed by APT, likely due to the formation of a carbon layer during electron imaging [34]. For APT studies, fresh specimens with Cr-coating from all four sides were prepared onto Si-micro posts [28]. All the Cr-coated specimens were transferred to the atom probe under ambient atmospheric conditions.

### 3.3 APT analysis of the Cr-coated specimens

**Figure 3 a-b-c** shows the XY and YZ projections of the 3D-reconstructions obtained from one of the coated NMC811 specimen, analyzed with a laser pulse energy of 5 pJ, 10 pJ and 15pJ, respectively. In the XY view, i.e. viewed along the Z-axis, all detected ions are displayed and the periphery is visibly filled with Cr-ions and Cr-O molecular ions. The sliced YZ view of the respective reconstruction shows the Li distributions in the specimen encapsulated by Cr-ions.

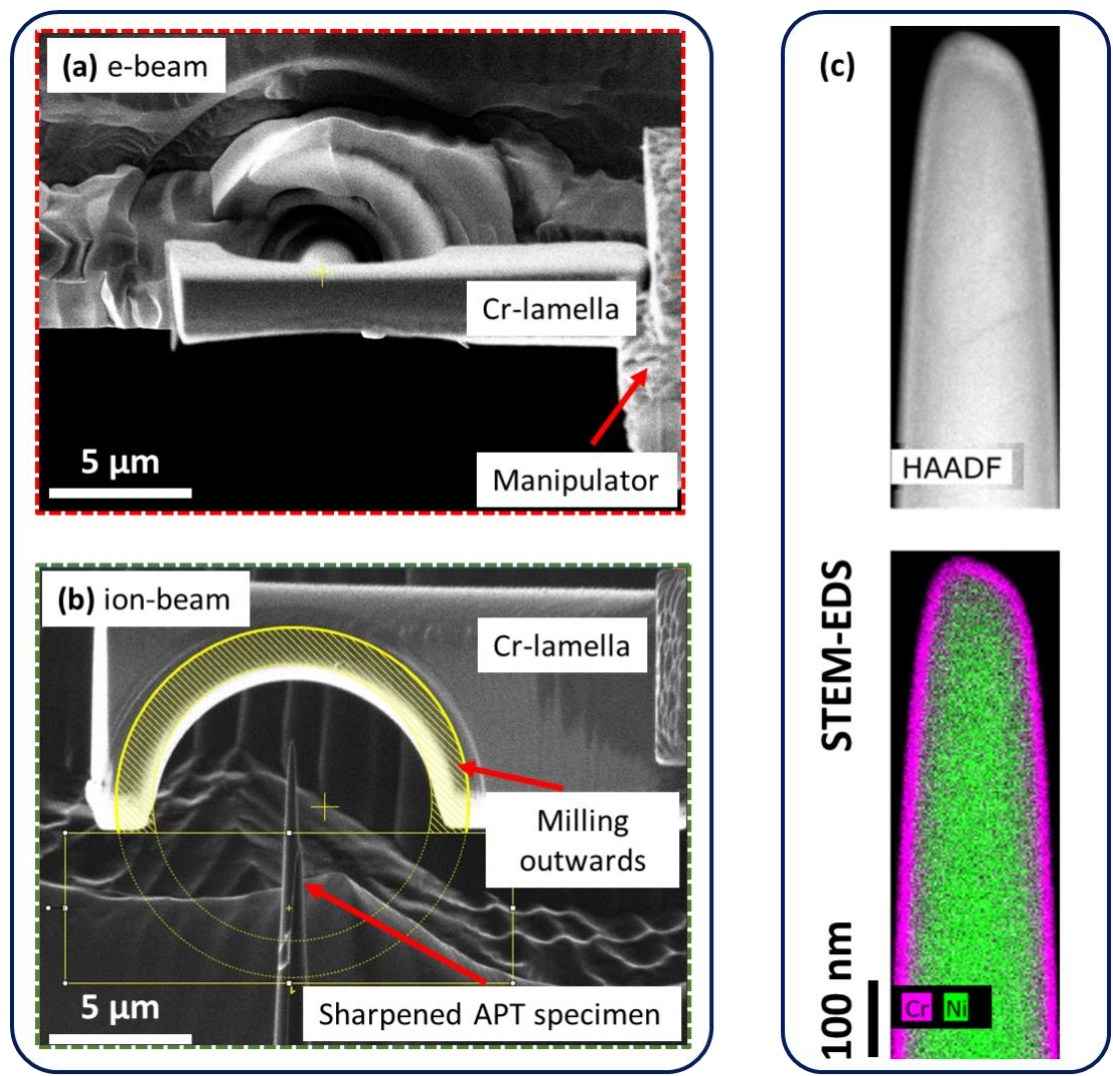

**Figure 2.** *(a) Electron beam and (b) ion-beam images of the aligned Cr-lamella on the sharpened NMC811 specimen, (c) STEM-HAADF image of the Cr-coated APT specimen of NMC811 sample. Coating was done from a single side at 30 kV, 80 pA for 30 secs. (d) STEM-EDS map showing a complete and thin presence of Cr on the APT specimen.*

The observation indicates that the Cr-coating enables homogeneous field evaporation of all constituents, particularly of Li-ions. For compositional analysis of the NMC811, i.e. excluding the coating itself, a cylindrical region-of-interest (ROI) 25 nm in diameter was placed at the center of the reconstructed data, parallel to the Z-axis. Similar ROI analysis were employed throughout our studies of coated specimens. The Ni, Li, and O compositions are plotted as a function of the depth in figure 3d, which was extracted from the 5 pJ dataset, as indicated by the dashed white arrow in figure 3a. The atomic compositions for the 10 pJ and 15 pJ is consistent with the 5pJ data.

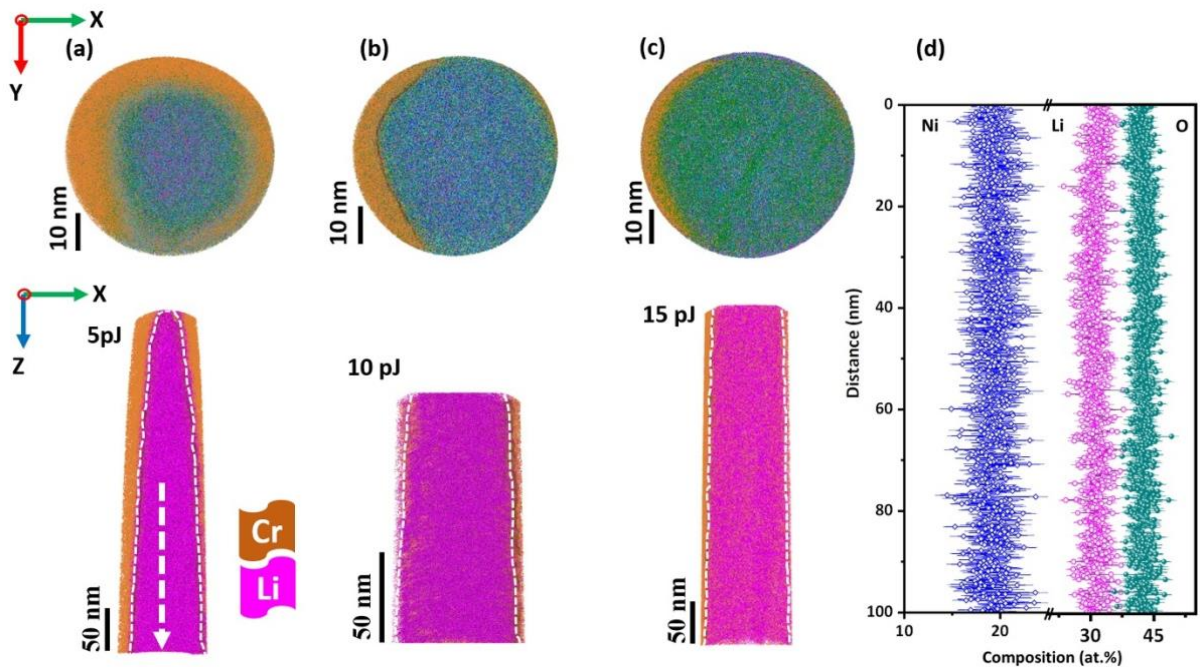

**Figure 3.** *Reconstructed atom map obtained from the atom probe measurements of the Cr-coated NMC811. Top of the image shows a slice of the dataset across the tip (XY plane) and bottom image along the tip (XZ plane) for the specimen ran at (a) 5pJ, (b) 10 pJ, (c) 15 pJ laser energy, (d) the compositional profile for the 5 pJ laser energy section measured along the tip using a cylinder (φ 25 x 25 nm²) in a region of interest such that Cr-coated is avoided, marked with white dashed line in 5pJ reconstruction.*

### 3.3.1 Mass-spectrum analysis and Spatial homogeneity

The mass spectra of the coated and uncoated NMC811 specimens acquired with the same parameters are plotted in Supplementary figure 1. The laser pulse energy is a relatively poor descriptor of the conditions at the specimen's surface during field evaporation. Typically this can be traced by using a charge state ratio (CSR) [36] as reported for metals [40,41], or molecular ion ratio (MIR) which can be more readily accessible in oxides, nitrides or hydrides for instance [35,42,43]. These incorporates analysis parameters that are critical for the field evaporation, such as specimen shape, base temperature, laser pulse energy, and detection rate. Schreiber et al. [35] in their study of Fe-oxides reported that $O_2^+/O^+$ was the most accurate descriptor for the electric field conditions, and it will be used in the following. The obtained ratio for $O_2^+/O^+$ is approx. 2.61 for both measurements.

For the Cr-coated samples, all the peaks corresponding to evaporated molecular ions (Ni-O, Mn-O and Co-O) are clearly distinguishable. As a note, there are additional peaks of Cr and Cr-O in the spectrum. Figure 4a compares the normalized mass spectra of the Cr-coated and uncoated obtained from the APT measurement at 5 pJ laser pulse energy, in the range of 4—35 Da. It illustrates the impact of Cr-coating on the mass resolution. The dissipation of the heat generated primarily at the end of the specimen [37,38] is accelerated by the conduction via the Cr coating, thus contributing to the reduction of the thermal tails. Some peaks previously hidden underneath the long thermal tails can now be observed and quantified. This outcome signifies an improved mass resolving power, resulting in more accurate chemical composition and sensitivity.

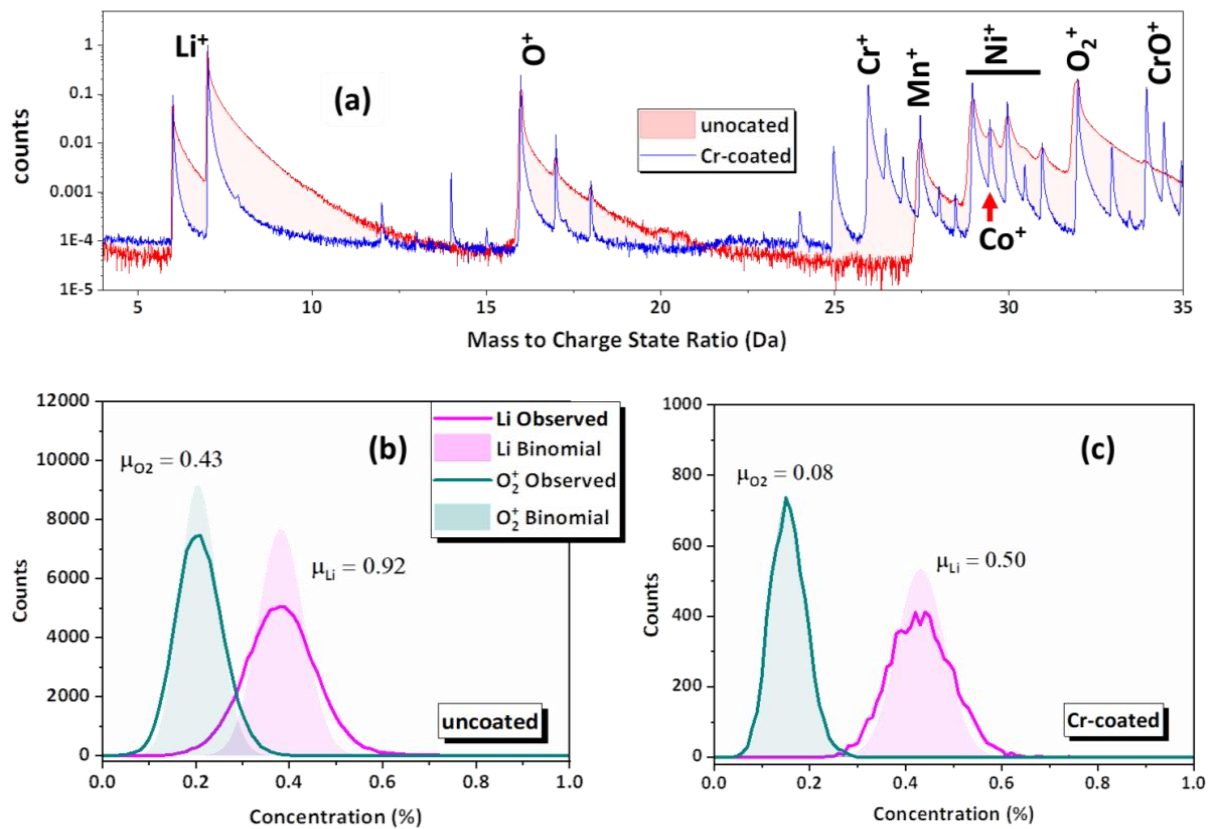

**Figure 4.** *Influence of the Cr-coating on the APT data measurement (a) Normalized and overlaid mass spectrum from uncoated and Cr-coated NMC811 sample, ran at 5pJ laser energy, measured under similar parameters (i.e. base temperature: 60K, detection rate: 0.5%, pulse frequency:*

*100 kHz), data is presented in a narrow range of 4-35 Da, illustrating an improvement in the mass resolving power for the coated sample. Frequency distribution analysis of the (b) uncoated and the (c) Cr-coated sample. Data is only presented for Li-ions and $O_2^+$ molecules. The deviation from the standard binomial distribution is assessed using Pearson coefficient (µ) as presented in the respected plots. Dashed lines are showing binomial distribution and solid lines are showing observed distributions.*

To understand the effect of the coating on the Li distribution, frequency distributions analyses [39] were performed, i.e. the region of interest is divided into bins, containing 100 ions, and the composition calculated and represented as a histogram as plotted in figure 4 b and c, along with the corresponding binomial distributions, that represents what would be expected from a completely random dispersion of atom in the analyzed volume [39]. The deviation from the random, binomial distribution can be quantified by using the Pearson coefficient (µ), which ranges from 0 for a random mixture to 1 for a fully separated distribution. In the case of the uncoated and Cr-coated samples, the frequency distribution analysis of Li⁺ ions yielded coefficients of 0.9 and 0.5 respectively. These results indicate that clustering has diminished with the application of the coating. A similar trend was observed for $O_2^+$ ions. Overall, the frequency distribution analysis strongly suggests that there are more pronounced solute agglomerations in the uncoated samples compared to those that have been Cr-coated.

### 3.3.3 Compositional variations with electric field conditions

A compositional analysis was performed across a wide range of laser pulse energies 5 – 50 pJ, while keeping all other parameters the same. Note that all specimens were prepared from a single secondary particle and coated with Cr. Details of the CSR vs. laser pulse energy for cations can be found in Supplementary figure 3, and as explained above, here we will use the ratio of $O_2^+$/O⁺ as a proxy for the intensity of the electrostatic field during the analysis [35].

Oxygen in laser pulsed APT of oxides can be underestimated [44,45]. This loss is attributed primarily to molecular dissociation forming neutral $O_2$ molecule either that cannot be detected, or that have a time-of-flight no longer proportional to the applied specimen voltage precluding their identification [44,46,47]. We have hence not included the O concentration in the analysis of the measured cation composition. Figure 5a displays the ratio Li/(Li+Ni+Co+Mn) plotted against $O_2^+$/O⁺ ratio. The expected ratio based on the materials stoichiometry is 0.5, and, remarkably, the observed ratio is close to this value across the entire range of electrostatic field.

To gain further insights into the evaporation behavior of the transition metal cations, the Ni/(Ni+Mn+Co) ratio is plotted in figure 5b as a function of the $O_2^+/O^+$, along with the atomic fraction of O obtained from the series of APT measurements. For Ni, the measured values remain steadily close to the expected ratio of 0.8, based on the stoichiometry, hovering around ~0.78 at lower $O_2^+/O^+$ ratios, i.e. lower laser pulse energies, but show a slight deviation at high $O_2^+/O^+$ ratios, corresponding to high laser pulse energy. For the anions, the expected value is 0.5, according to stoichiometry, and the measured ratio remain close to this expected value, especially with $O_2^+/O^+$ ratios below 4.

This systematic study of the cation and anion behavior as a function of the field conditions suggests that the Cr-coated samples can be analyzed across a wider range of laser pulse energies. Previous APT studies on such cathode materials had used lower laser energies [12,19,48], to mitigate delithiation. If compositional variations are minimal with lower laser pulse energies, Cr-coated specimens could be analyzed with higher laser pulse energies (lower electric fields), while maintaining a near-random Li-distributions, as tabulated in supplementary table 1, when compared to the data reported in figure 1. This is significant as a broader range of acceptable experimental conditions makes it easier to obtain large amounts of data, and that the data we provide herein can be used as a benchmark based upon which data can be compared.

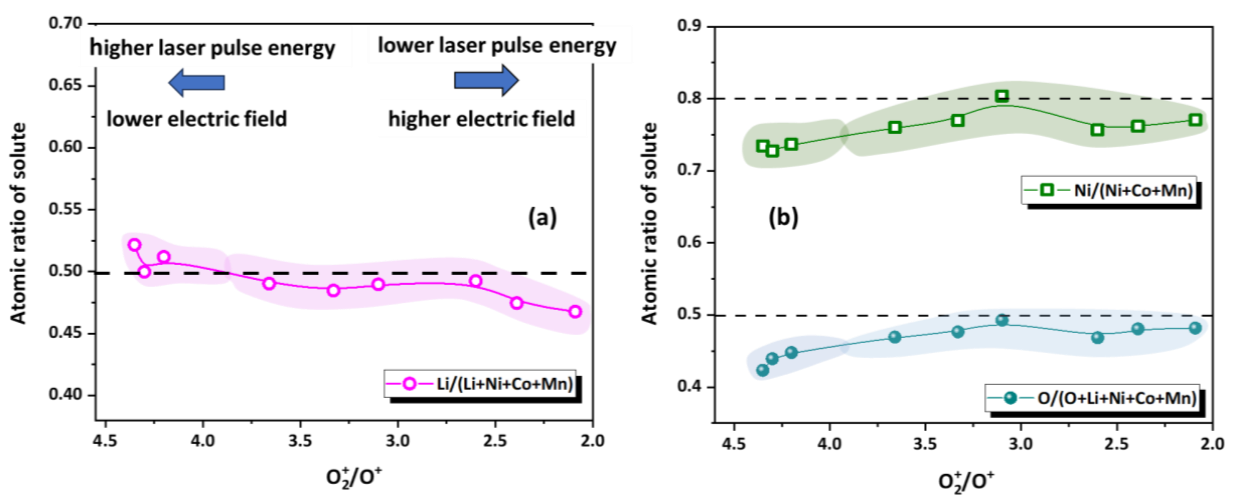

**Figure 5.** *Compositional quantification for NMC811 powder sample coated with Cr, plot of charge state-ratio of O ($O_2^+/O^+$) against the atomic ratio of the solutes (a) Li/(Li+Ni+Co+Mn), (b) Ni/(Ni+Co+Mn) and O/((Li+O+Ni+Co+Mn)). The expected values for the respective atomic ratio are*

*marked with horizontal dashed lines in the plots. All the measurements were done on the same secondary particles.*

### 3.3.4 Compositional in-homogeneity of NMC811 particles

Micro- and nano- heterogeneity in powder samples can arise during synthesizing bulk powder sample. This kind of heterogeneity is challenging to detect through bulk characterization techniques, but it can significantly impact the performance of cathodes in batteries. We employed APT to measure the composition across multiple secondary particles from the same batch of commercial powder. The compositional results obtained appear to consistently display a narrow range of variations, providing valuable and accurate way of measuring nano-scale chemistry of the battery materials such as NMC811, shown in figure 6.

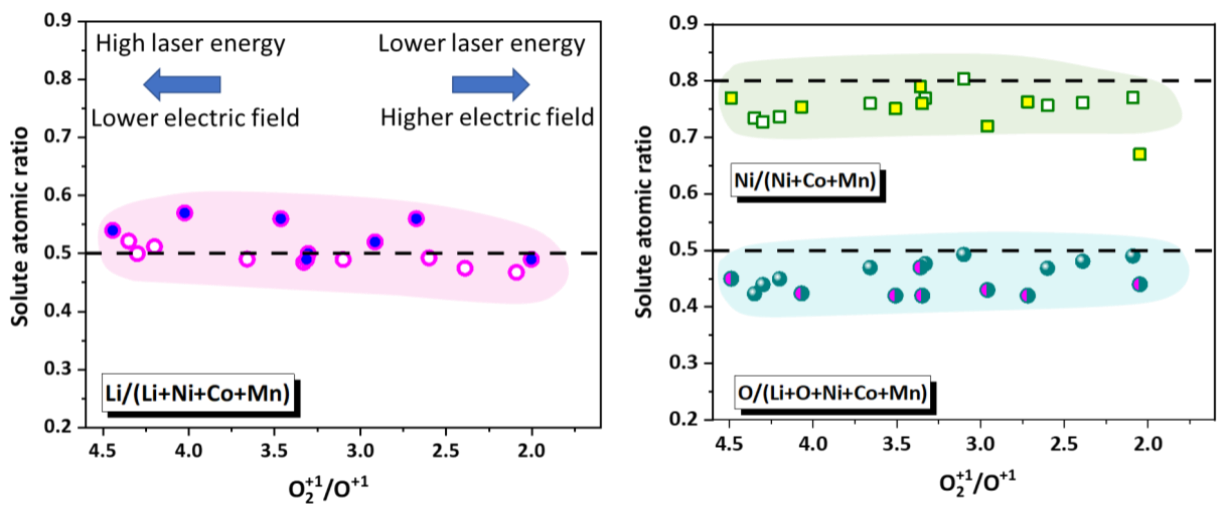

**Figure 6.** *Plots of solute atomic ratio for (a) Li/(Li+Ni+Co+Mn), (b) Ni/(Ni+Co+Mn) and O/(Li+O+Ni+Co+Mn) against $O_2^+/O^+$ ratios from multiple APT runs performed on the Cr- coated NMC811 samples. The open symbols are from data from figure 5 and the filled symbols are from independent analysis from separate particles with different laser energies. The expected values for the respective atomic ratio are marked with horizontal dashed lines in the plots.*

We have also encountered a region rich in Li- and Mn, likely to be an LMO particle within the NMC811. Figure 7a shows the APT reconstruction of one such Mn-rich particle. The obtained mass spectrum does not exhibit any peaks associated with Ni, Co and even Ni-O, Co-O molecular ions. The indexed mass spectrum is provided in the Supplementary figure 4. To understand the field evaporation behavior of these particles CSR values of Mn and O were plotted against the applied laser pulse energies, shown in figure 7b with black and yellow filled symbols. Following the post-ionization theory, at higher laser energies (low electric fields), the $Mn^{+2}/Mn^+$ should be lower, however it shows an opposite trend in the present samples. A similar behavior for $Mn^{+2}/Mn^+$ was observed in the Ni-rich NMC811 samples, as illustrated in Supplementary figure 3. There is a monotonous decrease in $Mn^{+2}/Mn^+$ ratio with decrease in laser energy, but no particular trend in the $O_2^+/O^+$ ratios, with only small variations, which may be explained by different field evaporation and dissociation trends for the different materials.

Specimens from a commercial LMO powder were prepared using the procedure previously used for NMC811 and were Cr-coated from all 4-sides. LMO is also an important and commercial used battery cathode material [45]. Subsequently, specimens were analyzed by APT using parameters similar to those used for NMC, with laser energies ranging from 1 pJ to 20 pJ. The obtained results of the CSR values of Mn and O are overlaid in figure 7b, with shaded regions. The $Mn^{+2}/Mn^+$ and $O_2^+/O^+$ ratios, show trends that are consistent with the LMO impurity particle reported above.

Figure 7c displays a plot of solute atomic ratios for Li and Mn as a function of $Mn^{+2}/Mn^+$, for both commercial LMO (shaded region) and impurity LMO obtained from NMC811 samples. In these plots, the $Mn^{+2}/Mn^+$ ratio is used since the $O_2^+/O^+$ ratio did not exhibit a clear trend. These fractions deviate significantly from the expected values, as marked with dashed lines, based on stoichiometry. This difference may be attributed to the suboptimal parameters for running Mn-rich oxides or, alternatively, it may raise questions about the evaporation behavior of Mn-ions in oxides. The elemental distribution is more homogeneous than in previous reports [14], including for Li, showcasing benefits of Cr-coating. The Li agglomerations, which was studied using Pearson coefficient (μ), consistently showed values in the range of 0.6—0.4. The data is presented in Supplementary table 1. This suggest that Cr-coating on LMO did improve the Li-ion field evaporation behavior, yet, further optimization of the APT parameters or trials with different metallic coatings (Ni, Ag, Ti) may be necessary to achieve optimal results.

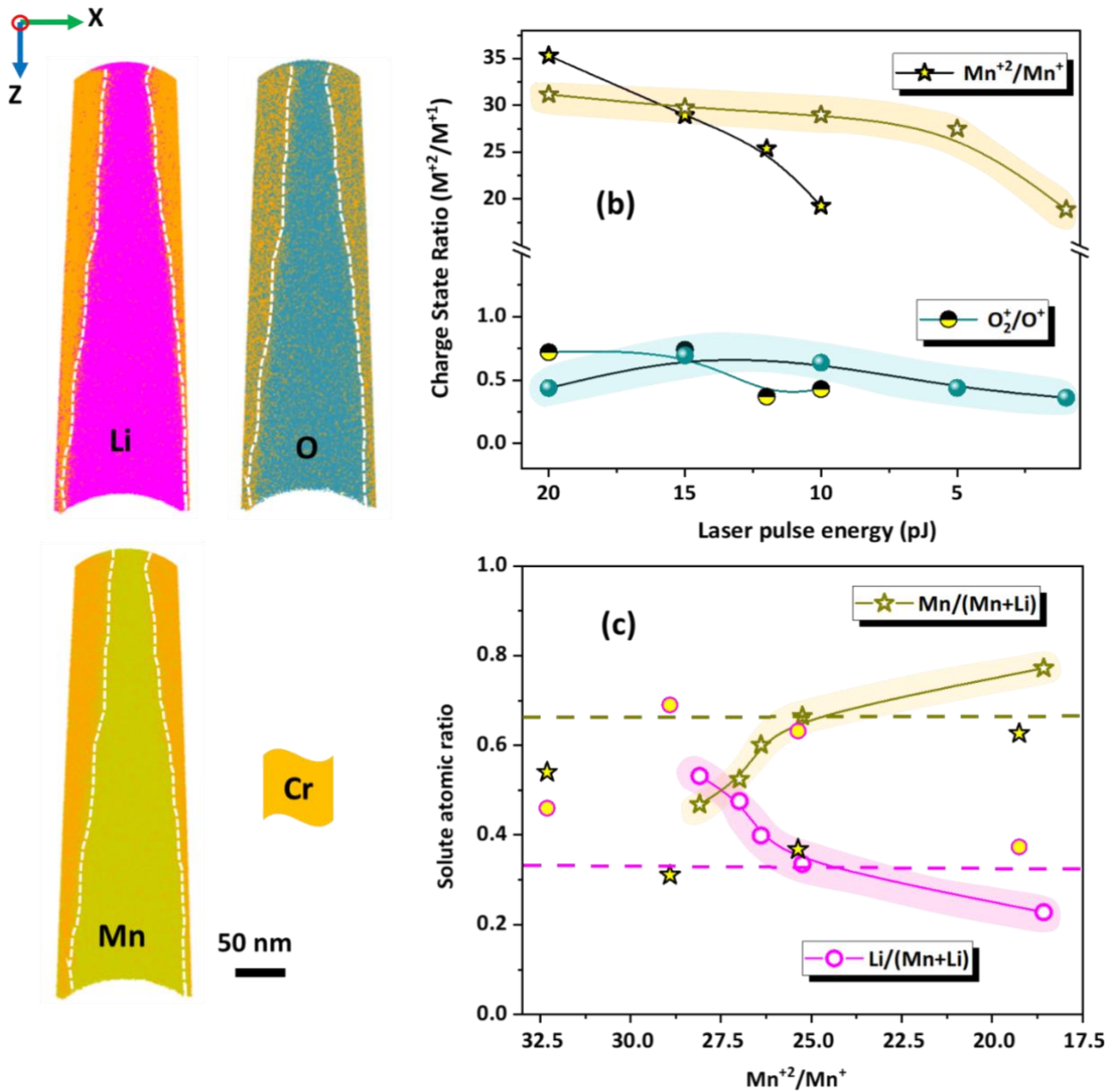

**Figure 7.** *Reconstructed dataset obtained from the atom probe measurements of the Cr-coated Mn-rich particle (a) sliced of the dataset along the tip (XZ plane) for the specimen ran at 20 pJ with base temperature of 60K, pulse frequency of 100kHz, and 0.5 % average evaporation rate, (b) plot of laser pulse energy against $Mn^{+2}/Mn^{+}$ and $O_2^+/O^+$ ratios obtained from the APT measurements for the Mn-rich samples (black-yellow symbols) and commercial $LiMn_2O_4$ powder (with shaded region) cathode samples, (c) Compositional quantification for the Mn-rich samples and commercial $LiMn_2O_4$ powder (with shaded region) coated with Cr, plot of charge state-ratio of Mn ($Mn^{+2}/Mn^{+}$) against the atomic ratio of the solutes (a) Mn/(Li+Mn), (b) Li/(Li+Mn), the expected values for the respective atomic ratio are marked with horizontal dashed lines in the plots.*

### 3.5 Perspectives

The high level of confidence in the results we provide now for the NMC811 can be effectively employed to understand the microstructural degradation of lithium-ion batteries (LIB) during successive charge and discharge cycles through APT studies. Additionally, the protocol presented in this work allows for the study of nanoscale compositional heterogeneities that can arise during material synthesis or during battery cycling. The nano-scale capability of APT showed that Mn-rich particles and off-stochiometric compositions are present in the commercial NMC811 samples, likely originating during the synthesis process. There are no reports on such inhomogeneity, largely because most characterization techniques study bulk composition [21,49–51] . While STEM-EDS or EELS studies can capture such inhomogeneities, they come with other complications [52]. APT, with its ability to capture nanoscale composition and structure information, has unveiled these issues that warrant attention during the synthesis.

### 4 Conclusion

Using *in-situ* redeposition through FIB milling of a metal target, we have devised a simple, cost-efficient and versatile method for applying metal shielding to APT specimens prepared from Li-containing materials for battery applications. As an example, we demonstrated Cr-coating on specimens from a batch of commercial $LiNi_{0.8}Co_{0.1}Mn_{0.1}O_2$ and $LiMn_2O_4$ particles. The coating provided an easy pathway for heat dissipation along with offering mechanical support to the specimens against the Maxwell stresses generated by the applied field, that are the main cause of failure of atom probe specimens [53,54]. The Cr-coating also led to notable improvements with regards to the homogeneity of the imaged Li distribution, the range of laser pulse energies that can be used to perform the analysis. The improved APT results also revealed solute inhomogeneity issues in bulk prepared samples, which is arises during synthesis.

### Acknowledgement

EVW and BG are grateful for funding from the ERC for the project SHINE (ERC-CoG) #771602. SHK and BG are grateful for funding from the German Research Foundation (DFG) through DIP Project No. 45080066.  EVW and BG are grateful to the DFG for funding through the 2020 Leibniz Award.  CJ is grateful for financial support from Alexander von Humboldt foundation. We thank Uwe Tezins, Christian Broß and Andreas Sturm for their support at the FIB and APT facilities at MPIE. Authors would like to thank colleagues at Imperial College London for providing the commercial NMC811 powder.

# Supplementary Information

**Facilitating the systematic nanoscale study of battery materials by atom probe tomography through in-situ metal coating**

Mahander P Singh[1], Eric V Woods[1], SeHo Kim[1,2], Chanwon Jung[1], Leonardo S Aota[1], Baptiste Gault[1,3]

1. Max-Planck-Institut für Eisenforschung GmbH, Max-Planck-Straße 1, 40237 Düsseldorf, Germany
2. *now at* Department of Materials Science and Engineering, Korea University, Seoul 02841, Republic of Korea
3. Department of Materials, Royal School of Mines, Imperial College London, Prince Consort Road, London SW7 2BP, UK

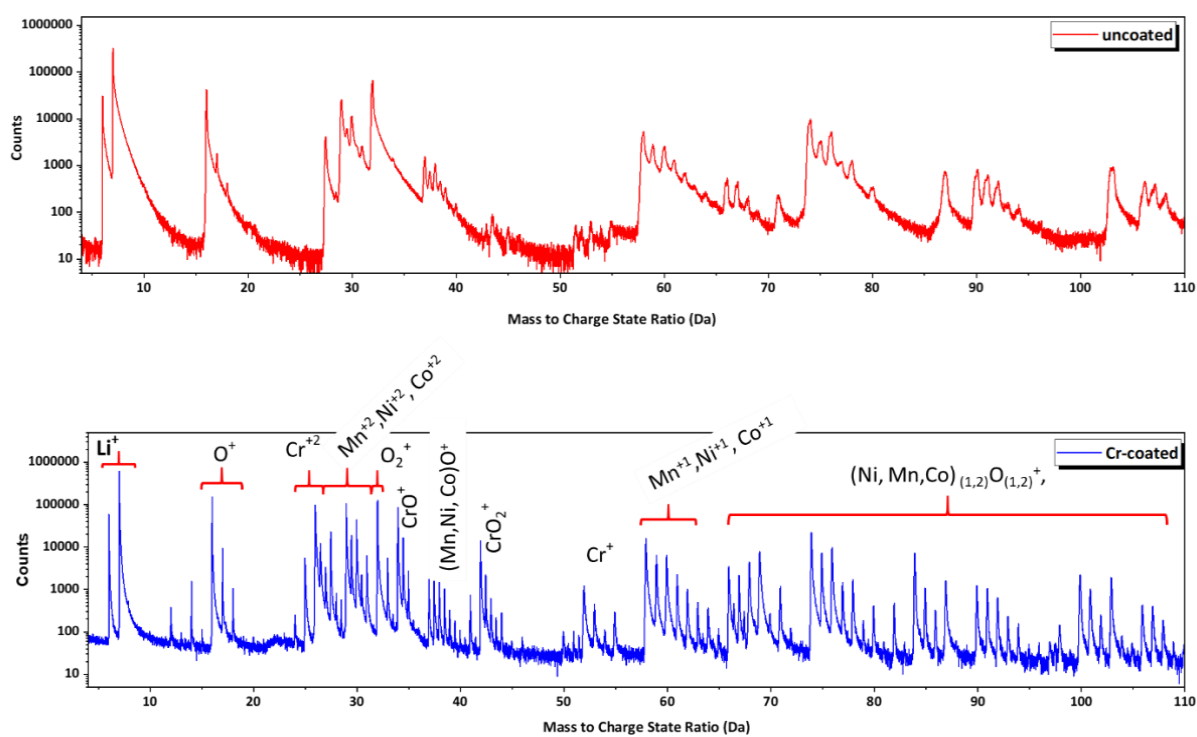

**Supplementary Figure 1:** mass-to-charge spectra comparison. Mass-to-charge spectra from (a) uncoated and (b) Cr-coated for commercial NMC811 (LiNi$_{0.8}$Co$_{0.1}$Mn$_{0.1}$O$_2$) samples analyzed using similar parameters (i.e. 5 pJ laser pulse energy, base temperature: 60K, detection rate: 0.5%, pulse frequency: 100 kHz). The ions observed to in the Cr-coated specimens are

Li$^+$, O$^+$, O$_2^+$, Mn$^+$, Mn$^{++}$, Ni$^+$, Ni$^{++}$ Co$^+$, Co$^{++}$

MnO$^+$, NiO$^+$, CoO$^+$, Mn$_2$O$^+$, Ni$_2$O$^+$, Co$_2$O$^+$, MnO$_2^+$, NiO$_2^+$, NiO$_2^+$, CoO$_2^+$, NiO$_3^+$, CoO$_3^+$, MnO$_4^+$

Cr$^+$, Cr$^{++}$, CrO$^+$, CrO$_2^+$, CrO$_3^+$,

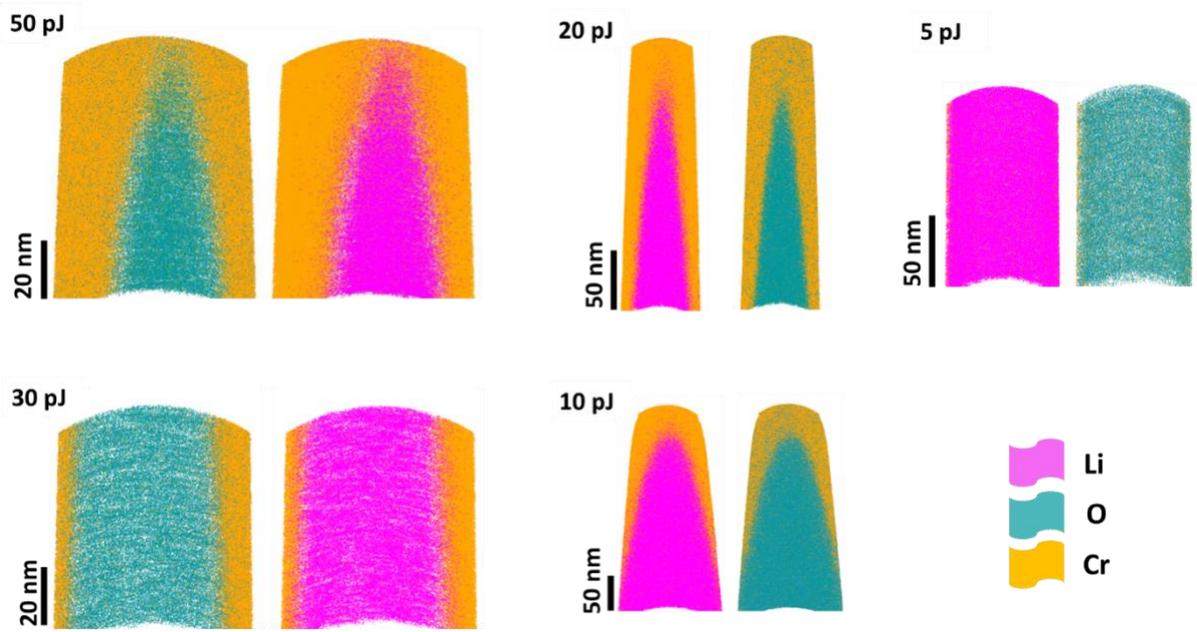

**Supplementary Figure 2:** 3D- reconstructions for the Cr-coated NMC811 samples analyzed at different laser pulse energies. Rest of the parameters were same.

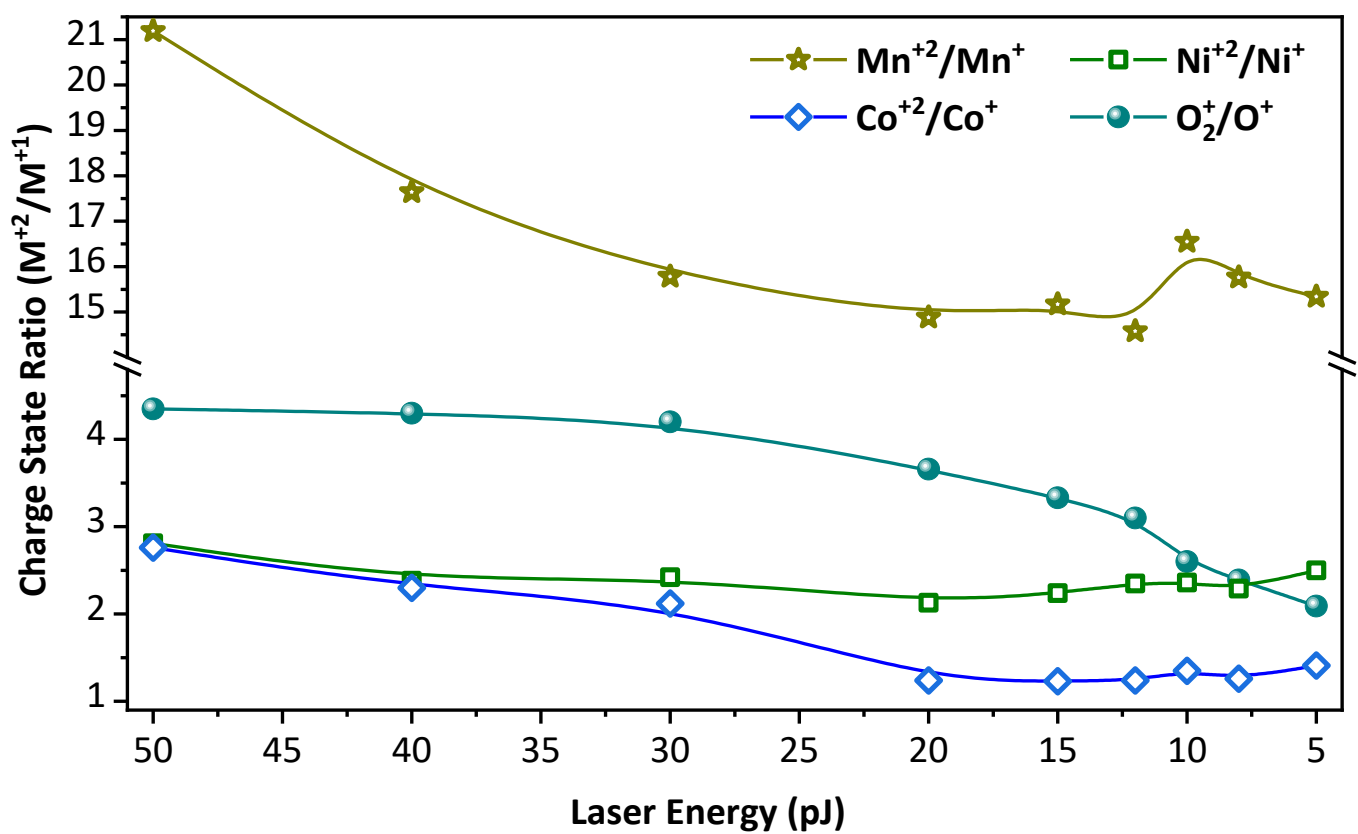

**Supplementary Figure 3:** *Charge state ratios of solute elements plotted against laser pulse energy.*

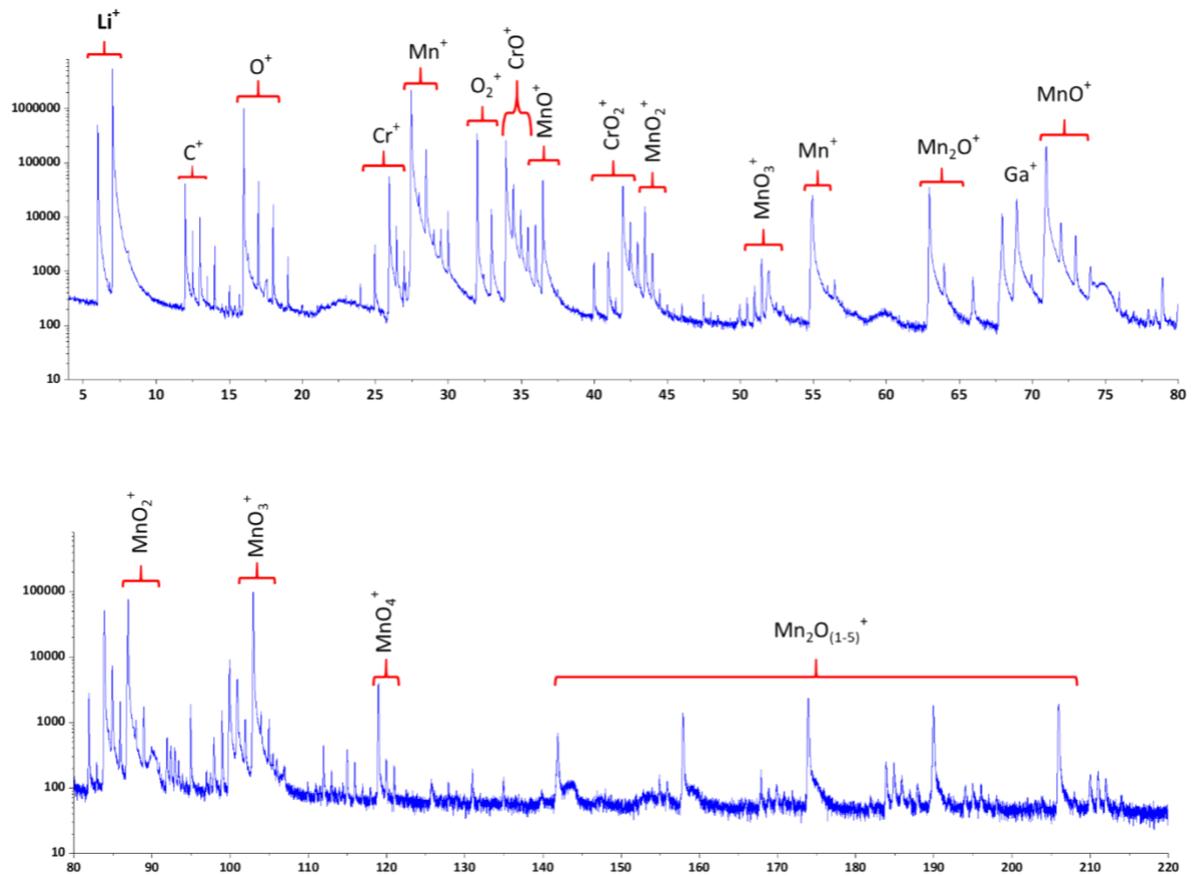

**Supplementary Figure 4:** mass-to-charge spectra for the Mn-rich particles form the commercial NMC811 samples analyzed using similar laser energy of 20 pJ, base temperature: 60K, detection rate: 0.5%, pulse frequency: 100 kHz)

| laser Pulse Energy (pJ) | NMC811 | Mn-rich particles from NMC811 | LiMn$_2$O$_4$ |
|---|---|---|---|
| 50 | 0.855 | | |
| 40 | 0.75 | | |
| 30 | 0.5163 | | |
| 20 | 0.501 | 0.561 | 0.621 |
| 15 | 0.521 | 0.5231 | 0.6518 |
| 12 | 0.5351 | 0.481 | |
| 10 | 0.5963 | 0.503 | 0.6145 |
| 8 | 0.498 | | 0.4783 |
| 5 | 0.4484 | | 0.3216 |

**Supplementary table 1:** *Pearson coefficient (µ) values for Li obtained from different APT runs and laser energies on NMC811, Mn-rich and LiMn$_2$O$_4$ samples.*